\documentclass[%
superscriptaddress,
showkeys,
preprint,
 amsmath,amssymb,
 aps,
prc
]{revtex4-1}

\usepackage{graphicx}
\usepackage{dcolumn}
\usepackage{bm}
\usepackage{booktabs}
\usepackage{soul, color, xcolor}
\usepackage{multirow}
\usepackage{tabularx}
\usepackage{float}
\usepackage{caption}

\usepackage{hyperref}

\usepackage{CJKutf8}
\begin{document}
\begin{CJK*}{UTF8}{gbsn} 
\preprint{APS/123-QED}

\title{Medium recoil mode of $\Delta$ production in single isobaric charge-exchange reactions}

\author{Xin Lei} %
\affiliation{Sino-French Institute of Nuclear Engineering and Technology, Sun Yat-sen University, Zhuhai 519082, China}
\author{Erxi Xiao}
\affiliation{Sino-French Institute of Nuclear Engineering and Technology, Sun Yat-sen University, Zhuhai 519082, China}
\author{Yingge Huang}
\affiliation{Sino-French Institute of Nuclear Engineering and Technology, Sun Yat-sen University, Zhuhai 519082, China}
\author{Yujie Feng}
\affiliation{Sino-French Institute of Nuclear Engineering and Technology, Sun Yat-sen University, Zhuhai 519082, China}
\author{Hui Wang}
\affiliation{Sino-French Institute of Nuclear Engineering and Technology, Sun Yat-sen University, Zhuhai 519082, China}
\author{Jiali Huang}
\affiliation{Sino-French Institute of Nuclear Engineering and Technology, Sun Yat-sen University, Zhuhai 519082, China}
\author{Fuchang Gu}
\affiliation{Sino-French Institute of Nuclear Engineering and Technology, Sun Yat-sen University, Zhuhai 519082, China}
\author{Long Zhu}
\affiliation{Sino-French Institute of Nuclear Engineering and Technology, Sun Yat-sen University, Zhuhai 519082, China}
\author{Jun Su}\email{sujun3@mail.sysu.edu.cn} %
\affiliation{Sino-French Institute of Nuclear Engineering and Technology, Sun Yat-sen University, Zhuhai 519082, China}
\affiliation{China Nuclear Data Center, China Institute of Atomic Energy, Beijing 102413, China}
\affiliation{Guangxi Key Laboratory of Nuclear Physics and Nuclear Technology, Guangxi Normal University, Guilin 541001, China}


\date{\today}

\begin{abstract}
The dynamic mechanisms underlying single charge-exchange reactions have been investigated using a theoretical framework that combines the Isospin-dependent Quantum Molecular Dynamics (IQMD) model with the statistical decay model GEMINI++.
Two distinct channels contribute to the single isobaric charge-exchange reaction: quasi-elastic channel, where neutron-proton scattering drives the charge-exchange, and inelastic channel, where the $\Delta$ particle is produced during the process.
In a referenced study [Phys.RevC 106.014618(2022)], experimental data have revealed that the inelastic channel accounts for approximately 50 percent of the single isobaric charge-exchange reaction.
However, our current model fails in reproducing the significant contribution of inelastic channel unless the novel medium recoil mode associated with $\Delta$ production is considered in the calculations.
Notably, this in-medium effect arising from inelastic nucleon-nucleon collisions is not yet incorporated into mainstream microscopic transport models.
The dynamical properties of protons and pions emitting in the single isobaric charge-exchange reactions are predicted. This exploration of in-medium effects adds a valuable dimension to our understanding of the intricate dynamics involved in single charge-exchange reactions.
\end{abstract}

\maketitle
\end{CJK*}

\section{\label{int}Introduction}
The baryon, consisting of three quarks bound by the strong interaction, serves as a fundamental building block in particle physics.
The study of the excitation spectra of baryon resonances has emerged as an intriguing topic in strong interaction physics \cite{friedman2007medium, hyodo2021qcd, hansen2019lattice, J2019Excitation}.
This exploration provides an effective tool for advancing our understanding of quantum chromodynamics \cite{brambilla2014qcd, gross202350} and investigating three-body forces \cite{krebs2018three}, given the three-quark structure inherent in baryons.

Among the common baryons, the proton and neutron play crucial roles in the components of atomic nuclei.
As the first excited mode of the nucleon, $\Delta$ resonances are produced in nuclear reactions at energies of hundreds of MeV/nucleon, subsequently decaying by emitting pions mesons \cite{pascalutsa2007electromagnetic}.
Pion production in these nuclear reactions serves as a valuable probe, offering insights into the high-density behavior of symmetry energy \cite{aoust2006pion, estee2021probing, gao2021pion, li2002probing}.
Symmetry energy is a key input for studying neutron stars \cite{li2018competition}.
Additionally, concerns arise regarding the potential influence of $\Delta$ resonances on neutron star properties, including mass and radii \cite{motta2020delta, cai2015critical,malfatti2020delta}.

The properties of the $\Delta$ resonance in nuclear environment exhibit variations comparing with those in the free space.
These in-medium effects have been proven in various experiments, and are crucial to understanding the nuclear reaction dynamics.
Observable downward energy shifts of approximately 70 MeV in the $\Delta$ peak for heavier targets compared to protons suggest the influence of the nuclear medium \cite{mukhin1995delta}.
The $\Delta$ mass shows a decreasing trend with increasing multiplicity have been observed in p+C collisions at 0.8 GeV \cite{trzaska1991excitation}.
Notably, the width of $\Delta$ resonances in hadron-nucleus collisions at 4.2 GeV/c is smaller compared to free nucleon collisions \cite{olimov2021delta, olimov2023production}, potentially attributed to the effect of Pauli blocking \cite{lenske2018hyperons}. Furthermore, the lifetime of $\Delta$ is reduced compared to that in free space, owing to interactions between $\Delta$ resonances and surrounding nucleons \cite{reichert2019delta}. The thresholds of the NN$\to$N$\Delta$ reaction in the medium differ from those in free space due to mean-field potentials \cite{song2015modifications}, with more detailed effects discussed in Ref. \cite{zhang2017medium}.

In microscopic transport models for nuclear reactions, the $\Delta$ and pion productions are performed via the cascade mode \cite{vidana2016excitation, wolter2022transport}, where the related channels include NN$\to$N$\Delta$, N$\Delta$$\to$ NN, $\Delta$$\to$$\pi$N, and $\pi$N $\to$$\Delta$ \cite{ramachandran1997delta, liu2021insights, Xie_2019}.
Typically, the empirical formulas of cross sections and decay with fitted to experimental data in free space are used as inputs \cite{DANIELEWICZ1991712,ono2019comparison}.
Incorporating in-medium effects into transport models have proven to be valuable, and several factors have been fruitful endeavors, such as pion and $\Delta$ potentials \cite{xu2010isospin, feng2017preequilibrium, zhang2018effects}, threshold effects \cite{song2015modifications}, nucleon correlations \cite{li2015effects}.

The single isobaric charge-exchange reaction has provided a focused and direct approach for studying the nucleon-nucleon interaction in nuclear medium \cite{ableev1984excitation, ableev1987p, ableev1988charge}.
Earlier study of $^{129}$Xe + $^{27}$Al at 790 MeV/nucleon revealed that, the formation of $\Delta$ in nucleon-nucleon collisions has an impact on the cross section of charge-exchange production $^{129}$Cs \cite{summerer1995charge}.
The subsequent investigations of $^{208}$Pb + $^{1,2}$H collisions at 1 GeV/nucleon showcased two distinct contributions for charge-exchange cross sections: one is the quasi-elastic process and the other is inelastic process related to the $\Delta$ production \cite{kelic2004isotopic}.
Using the missing-energy spectra of the single isobaric charge-exchange productions, the cross sections of $^{112}$Sn and $^{124}$Sn isotopes at 1 GeV/nucleon are separated into two components: quasi-elastic channel and inelastic channel \cite{rodriguez2022systematic}.
These reactions provide a unique perspective to understand the $\Delta$ resonances in nuclear medium.

This study focuses on single isobaric charge-exchange reactions occurring in $^{112}$Sn + p and $^{136}$Xe + p collisions at 1 GeV/nucleon.
In these collisions, the projectiles $^{112}$Sn ($^{136}$Xe) undergo one-charge change process, leading to the production of In or Sb (I or Cs) isotopes.
At the energy of 1 GeV/nucleon, both elastic and inelastic nucleon-nucleon collisions play roles in the single charge-exchange process.
This work specifically focuses on the latter channel, exploring the dynamics associated with the production of $\Delta$ particles in the context of single charge-exchange reactions.
The paper is organized as follows.
In Sec. \ref{model}, the theoretical framework is described.
In Sec. \ref{results}, both the results and discussions are presented.
Finally, Sec. \ref{summary} presents the conclusions

\section{\label{model}Theoretical framework}
\subsection{\label{sub1} Isospin-dependent quantum molecular dynamics model}
The Quantum Molecular Dynamics (QMD) model is widely used in the description of heavy-ion collisions which can dated back to 1988 \cite{aichelin1988quantum}, employing the event-by-event simulation method.
The isospin-dependent Quantum Molecular Dynamics model (IQMD) that includes isospin-dependent interaction based on the original QMD is applied in this work \cite{1989Quantum}.
The version of the IQMD code used in this paper is IQMD-BNU (Beijing Normal University), which has been introduced and compared to other versions within the Transport Model Evaluation Project \cite{wolter2022transport}.

In the IQMD model, the single nucleon is treated as the Gaussian wave packet within the coherent state:
     \begin{equation}	
          \phi _\mathit{i} (\mathbf{r},\mathit{t}  )= \frac{1}{(2\pi \mathit{L})^{3/4}} e^{-\frac{[\mathbf{r}-\mathbf{r_\mathit{i}(\mathit{t} ) }  ]^2}{4L} }e^{\frac{\mathit{i}\mathbf{r}\cdot \mathbf{p}_\mathit{i}(\mathit{t} )    }{\hbar } },
     \end{equation}
where $\mathbf{r}_{i}$ and $\mathbf{p}_{i}$ represent respectively the average position and momentum of the $i$-th nucleon, and $L$ is corresponded to the square of the Gaussian wave packet, which is taken 2.0 fm$^{2}$ in our work.
The nucleus system is described by the N-body wave function, which can be given as the direct product of wave function of single nucleon.
The wave function can be transformed into a distribution of phase-space density function by applying the Wigner transformation, which can be expressed as follows:
\begin{equation}
f(\mathbf{r},\mathbf{p},t) = \sum_{\mathit{i=1}}^{\mathit{N}} \frac{1}{{(\pi \hbar )}^3}e^{-\frac{[\mathbf{r}-\mathbf{r_\mathit{i}(\mathit{t} ) }  ]^2}{2L} }e^{-\frac{[\mathbf{p}-\mathbf{p}_\mathit{i}(t)]^2 \cdot 2\mathit{L}   }{\hbar ^2} },
\end{equation}
The density distribution function of N-body system can be obtained by integrating of the phase-space density function $f(\vec{r},\vec{p},t)$, the functions are as follows:
\begin{equation}
\rho(\boldsymbol{r})= \frac1{(2\pi L)^{3/2}}\sum_{i=1}^N\mathrm{e}^{-[r-r_i(t)]^2/2L}
\end{equation}
The time evolution of the nucleons in the self-consistently generated mean field is determined by the Hamiltonian equations of motion,
\begin{equation}
 \dot{\mathbf{r} }_\mathit{i} = \nabla _{\mathbf{p}_i }\mathit{H},  \dot{\mathbf{p}}_i = -\nabla _{\mathbf{r}_i}\mathit{H}.
\end{equation}
The Hamiltonian consists of three parts: the kinetic energy, Coulomb potential energy, and nuclear potential energy.
The detailed expression is as follows:
\begin{equation}
\mathit{H}=\mathit{T} +  U_{coul} +  \int V\left [ \rho \left ( \mathbf{ r} \right )  \right ] \mathrm{d}\mathbf{r}
 \end{equation}
The nuclear potential energy of the asymmetric nuclear matter with density $\rho$ and asymmetry $\delta$ is given by
\begin{equation}
\begin{aligned}
V(\rho, \delta) = & \frac{\alpha}{2} \frac{\rho^2}{\rho_0} + \frac{\beta}{\gamma+1} \frac{\rho^{\gamma+1}}{\rho_0^{\gamma}} + \frac{C_{sp}}{2}(\frac{\rho}{\rho_{0}})^{\gamma_{i}} \rho \delta ^{2},
    \label{V}
  \end{aligned}
  \end{equation}
where $\rho_0$ is the normal density. 
The parameters $\alpha$, $\beta$, $\gamma$, $C_{sp}$, and $\gamma_{i}$ are temperature-independent terms.
These parameters used in this work are $\alpha$ = -356.00 MeV, $\beta$ = 303.00 MeV, $\gamma$ = 7/6, $C_{sp}$ =38.06 MeV, and $\gamma_i$ = 0.75.

The IQMD model incorporates the nucleon-nucleon (NN) collision to account for both the short-range repulsive residual interaction and the stochastic changes in the phase-space distribution, which is given as:
\begin{equation}
(\frac{\mathrm{d} \sigma }{\mathrm{d} \Omega })_\mathit{i}   = \sigma _\mathit{i}^{free}\mathit{f_\mathit{i}^ {angl} }\mathit{f}_\mathit{i} ^{med} ,
\end{equation}
where $\sigma^{free}$, $\mathit{f^{angl}}$, and $\mathit{f}^{med}$ represent the cross section of NN collisions in the free space, the angular distribution, and the in-medium corrections, respectively.
The subscript $\mathit{i}$ is related to distinguish the different channels of the NN collisions, i.e., the elastic proton-proton scatterings ($\mathit{i}$ = pp), elastic neutron-proton scatterings ($\mathit{i}$ = np), elastic neutron-neutron scatterings ($\mathit{i}$ = nn), and inelastic nucleon-nucleon collisions.
The parameterization of $\sigma^{free}$ and $\mathit{f^{angl}}$ are isospin-independent are taken from\cite{CUGNON199621557}.
The $\mathit{f}^{med}$ of elastic scatterings is given by \cite{coupland2011influence58}
\begin{equation}
\begin{aligned}
\mathit{f_\mathit{el}^ {med} } &=\sigma _\mathit{0}/\sigma^{free}\tanh(\sigma^{free}/\sigma _\mathit{0}),\\
\sigma _\mathit{0}&=0.85\rho ^{-2/3} .
\end{aligned}
\end{equation}

To compensate for the fermionic feature, the phase-space density constraint (PSDC) \cite{papa2001constrained} method and the Pauli blocking are applied in the IQMD model.
The phase-space distribution $\mathit{ f_{i}} (\mathbf{x_{i} },\mathbf{p_{i} })$ is gone for checking the Pauli blocking, where $\mathbf{x_{i} }$ and $\mathbf{p_{i}}$ represent the position of the scattering position and final state momentum of the particles, respectively.
The binary $NN$ collisions are allowed with the probability $(1-f_{i}^{'} )(1-f_{j}^{'})$.

Based on the PSDC, the probability of phase-space occupation $\overline{\mathit{f}_i}$ is calculated by performing the integration on a hypercube of volume $\mathit{h}^3$ in the phase space centered around the $\mathit{i}$th nucleon.
\begin{equation}
\bar{f}_i =0.621+\sum_{j\ne 1}^{N} \frac{\delta _{\tau _j,\tau _i}}{2} \int_{h^3}\frac{1}{\pi ^3\hbar ^3}e^{-\frac{(r_j-r_i)^2}{2L}- \frac{(p_j-p_i)^2}{\hbar ^2/2L}}d^3rd^3p.
\label{ith_possibility}
\end{equation}
Here 0.621 is the contribution itself and $\tau _i$ means the isospin degree of freedom.
The phase-space occupation $\overline{\mathit{f}_i}$ for each nucleon is checked at each time step.
If the value of phase-space occupation $\overline{\mathit{f}_i}$ is greater than ${\mathit k_{fcon}}$, the momentum of the $\mathit{i}$th nucleon is changed randomly by the many-body elastic scattering.
Meanwhile, the Pauli blocking in the binary NN collisions is modified.
Only if $\overline{\mathit{f}_i}$ and $\overline{\mathit{f}_j}$ at the final states are both less than 1, the result of many-body elastic scattering is accepted.

During the evolution process, the hot nuclei is differentiated by the minimum spanning tree (MST) algorithm.
The relative distance of coordinate is expressed as $\mid r_i-r_j \mid \le R_0$, while the momentum of nucleons is $\mid p_i-p_j \mid \le P_0$ in a hot system.
The $R_0$ and $P_0$ are equal to 3.5 fm and 250 MeV/c, respectively.
The excitation energy for each nucleon of pre-fragments formed in the hot system can be written as
\begin{equation}
        E^*=\frac{ {\textstyle \sum_{i}U_i}+ \sum_{i}\frac{(p_i-p_f)^2}{2m}-B(Z_f,A_f) }{A_f}.
 	\label{Ex}
\end{equation}
Here $U_i$ and $p_i$ are the single-particle potential and momentum of the $i$th nucleon; $p_f$, $Z_f$, and $A_f$ are the average momentum, charge number, and mass number for each nucleon of the fragment.
The binding energy of a nucleus is given as B($Z_f$,$A_f$), with charge number $Z_f$ and mass number $A_f$.
The summation is for the nucleons belonging to the same hot system.
The MST algorithm is performed at each time step, so the projectile spectator can be recognized.
The more details are shown in the Ref. \cite{Su2018Dynamical}.

\subsection{\label{sec:GEMINI++}GEMINI++}
The dynamical process takes the responsibility for describing the excitation stage of the nucleus.
In order to reduce the effect of virtual nucleons emission (a few nucleons will be evaporated even if the simulation of a nucleus is on the ground state) on the number of pre-fragments, when the excitation energies of the two heaviest pre-fragments are less than a certain value $E_{stop}$ ($E_{stop}$ = 3 MeV/nucleon \cite{BORDERIE2008551}), the process of the IQMD model will be stopped and the statistical-model code GEMINI++ will be switched on.
GEMINI++  developed by R. J. Charity \cite{charity2010systematic}, optimizes the overestimation of the prediction of fission mass distribution in heavy systems based on GEMINI \cite{charity1988systematics}.
The information of Z, A, $E^{*}$, and spin J formed of the compound nuclei formed in the dynamic process are taken as the input parameters in the model.

A sequential binary-decay is used to describe the de-excitation process of the compound nucleus which would undergo evaporation, fission and possible other decay modes to reach the ground state.
This process will stop until particle emission becomes energetically forbidden or improbable due to competition with gamma-ray emission.
Emission of light nuclei and proton/neutron is given by the Hauser-Feshbach formalism \cite{hauser1952inelastic}
The total fission yield is obtained from the Bohr-Wheeler formalism, more details are shown in the \cite{mancusi2010unified}.
The width of the fission-fragment mass distribution is expressed by the formula of Rusanov et al \cite{rusanov1997features,1980Deep}, which is applied to choose the mass division from the final scission temperature.
The secondary productions  obtained from the evaporation or fission processes can continue to undergo decay,  they will be persisted until particle emission becomes energetically prohibited or unfeasible due to competition with $\gamma$-ray emission.
In this study, the ratio $a_f/a_n$ is 1.036, the transient time (fission delay) is 1 zs, and the fission barrier is 1 MeV.
A more detailed description of GEMINI++ is presented in the Ref. \cite{charity2010systematic,mancusi2010unified}.

\section{\label{results}Results and discussion}

\subsection{Single charge-exchange reaction and its inelastic contribution}

\begin{figure}
\centering
\includegraphics[width=8.6cm,angle=0]{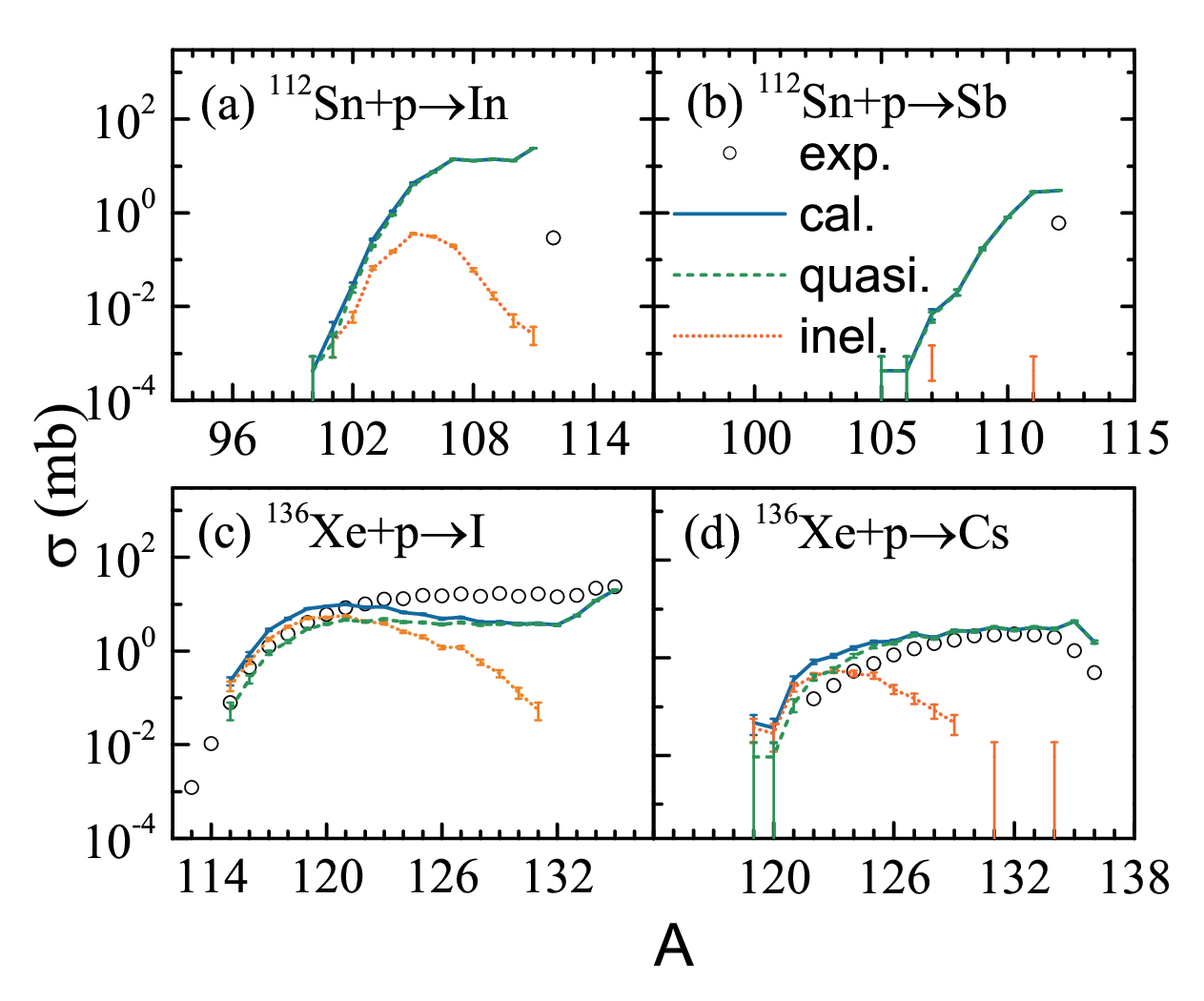}
\caption{ Production cross sections of (a) In and (b) Sb isotopes produced in the $^{112}$Sn + p collision, as well as (c) I and (d) Cs isotopes produced in the $^{136}$Xe + p collision at 1 GeV/nucleon.
Open circles are data taken from Ref. \cite{rodriguez2022systematic, napolitani2007measurement}.
The (blue) solid lines are the calculations of total cross sections.
The (orange) dotted and (green) dashed lines represent the inelastic, quasi-elastic cross sections.
}
\label{sigma_cas}
\end{figure}

The single charge-exchange reaction is a type of nuclear reaction that involves a one-charge change process between the projectile and the target.
In the context of the $^{112}$Sn + p collision, the production of indium (In) isotopes results from the removal of one charge from the projectile, whereas the production of antimony (Sb) isotopes is related to the addition of one charge to the projectile.
Similarly, in the $^{136}$Xe + p collision, the production of iodine (I) and cesium (Cs) isotopes is due to one-charge change processes.

To study the mechanism of one-charge changing, collisions involving $^{112}$Sn and $^{136}$Xe projectiles with protons at 1 GeV/nucleon were simulated using the theoretical framework that combines the Isospin-dependent Quantum Molecular Dynamics (IQMD) model with the statistical decay model GEMINI++.
The production cross sections of In and Sb isotopes from the $^{112}$Sn + p collision, as well as I and Cs isotopes from the $^{136}$Xe + p collision, were calculated and compared with available experimental data.
Figure \ref{sigma_cas} presents the calculated production cross sections (full curves) alongside the experimental data (open circles).
For the $^{112}$Sn + p collision, available data only include the production cross sections for $^{112}$In and $^{112}$Sb \cite{rodriguez2022systematic}.
The model significantly overestimates these cross sections.
In the case of the $^{136}$Xe + p collision, the data are taken from Ref. \cite{napolitani2007measurement}.
Here, the calculations are consistent with the data for $^{127-133}$Cs isotopes.
However, the model underestimates the production cross sections for $^{123-134}$I and generally overestimates the production cross sections for neutron-deficient isotopes of both I and Cs.
These discrepancies suggest that while the IQMD + GEMINI++ framework provides a reasonable overall description of the production cross sections for certain isotopes, it may require further refinement to accurately predict the cross sections of all isotopes, particularly those that are neutron-deficient.
This could involve improving the modeling of the nuclear reaction mechanisms or incorporating additional factors that influence the charge-changing processes.

The dynamic simulations indicate that both quasi-elastic and inelastic channels contribute to the production of isotopes displayed in Fig. \ref{sigma_cas}.
In the quasi-elastic channels, only nucleons are involved in the reaction dynamics.
These channels typically involve direct interactions where nucleons are scattered or transferred without the production of intermediate particles like pions or $\Delta$ particles.
In the inelastic channels, a $\Delta$ particle is produced and subsequently decays during the reaction.
This mechanism can be exemplified by the production of $^{112}$Sb in the $^{112}$Sn + p collision, which can be described as:
   \begin{equation}
   \begin{aligned}
   ^{112}\text{Sn} + p \rightarrow ^{112}_{\Delta^{0}}\text{Sn} + p, \\
   ^{112}_{\Delta^{0}}\text{Sn} \rightarrow ^{112}\text{Sb} + \pi^-.
   \end{aligned}
   \label{sb2}
   \end{equation}
In the initial interaction, one of the neutrons in the projectile $^{112}$Sn collides with a proton from the target, leading to the production of a $\Delta^0$ particle within the $^{112}$Sn nucleus.
The $\Delta^0$ particle, produced in the first step, decays into a proton and a $\pi^-$ meson.
The decay of $\Delta^0$ in the $^{112}$Sn nucleus results in the conversion of a neutron to a proton, producing the $^{112}$Sb nucleus alongside a $\pi^-$ meson and a proton.
The reaction steps outlined in Eq. (\ref{sb2}) highlight the complex dynamics involved in the inelastic channel, where intermediate particles play a crucial role in the isotope production process. These mechanisms contrast with the quasi-elastic channels, which are simpler and involve only nucleon interactions.

Understanding the contributions of both quasi-elastic and inelastic channels is essential for accurately modeling and predicting the production cross sections of various isotopes in nuclear reactions.
The production cross sections in Fig. \ref{sigma_cas} are divided into quasi-elastic and inelastic contributions, shown as (orange) dotted and (green) dashed curves respectively.
The calculations indicate that the quasi-elastic and inelastic cross sections are comparable on the neutron-deficient side.
As the mass number increases, the percentage of inelastic cross sections decreases rapidly.
For mass numbers close to that of the projectile, the inelastic cross sections are almost negligible.
The discrepancies observed in the calculated cross sections for isotopes in Figure \ref{sigma_cas} could be attributed to the complex interplay between these channels and the limitations of current theoretical models in capturing all aspects of the reaction dynamics.

\begin{figure*}
  \centering
  \includegraphics[width=14cm,angle=0]{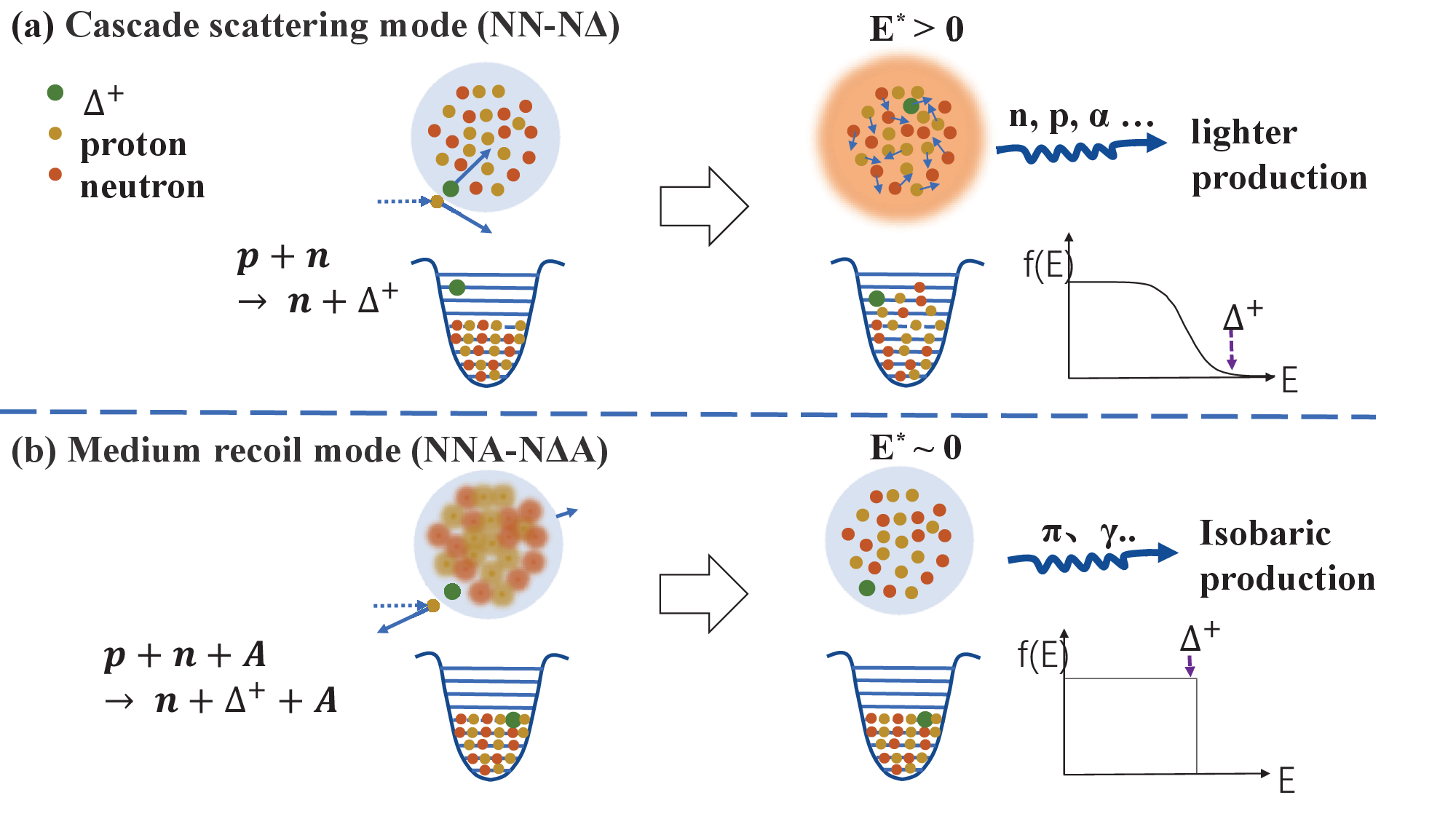}
  \caption{ Two energy transfer ways in the $\Delta$ production process: (a) Cascade scattering mode and (b) Medium recoil mode.
N refers to proton or neutron. Solid circles in dark yellow (orange) represent protons (neutrons), and the $\Delta$ particle is painted as green solid circle.
Target is depicted by light blue and orange circle for the ground and excited states, respectively.
Position of the $\Delta$ particle on the Fermi distribution function graph in two modes is illustrated. }
  \label{cas_med}
\end{figure*}

The $\Delta$ dynamics in the model is analyzed to better understand the inelastic channel.
In the traditional IQMD model, the $\Delta$ particle is generated through nucleon-nucleon cascade collisions and subsequently propagates in the mean field until it decays into nucleons and pions. Figure \ref{cas_med} (a) illustrates this $\Delta$ dynamics, labeled as the 'cascade scattering mode'.

Consider a scenario where a heavy ion collides with a proton at GeV/nucleon energy.
One of the neutrons in the projectile collides with a target proton and transforms into a $\Delta$ particle.
Due to conservation of momentum, the $\Delta$ particle recoils and is energetically boosted above the Fermi surface.
During its subsequent propagation, the $\Delta$ particle slows down through elastic scattering with other nucleons in the projectile.
Consequently, some nucleons in the projectile are raised above the Fermi surface.
Following these processes, the excitation energy of the projectile-like fragment becomes significant, leading to the vaporization of several neutrons in the secondary decay.
This results in the production of projectile-like fragments with a lower mass number compared to that of the original projectile.

This cascade scattering mode of the inelastic channel provides a reasonable explanation for the calculations of the inelastic cross sections shown in Fig. \ref{sigma_cas}, where the inelastic cross sections are almost negligible in the mass region near the projectile.
However, significant contributions from the inelastic channel to the production cross sections of $^{112}$In and $^{112}$Sb in the $^{112}$Sn + p collision at 1 GeV/nucleon have been observed \cite{rodriguez2022systematic}.
The model's calculations within the cascade scattering mode fail to reproduce these significant contributions from the inelastic channel.
This discrepancy has motivated efforts to enhance the $\Delta$ dynamics in the IQMD model.

The cascade scattering mode for $\Delta$ dynamics in the IQMD model is replaced by the medium recoil mode, as illustrated in Fig. \ref{cas_med} (b).
In this mode, the inelastic nucleon-nucleon collision that generates the $\Delta$ particle occurs within the nuclear medium, implying an in-medium effect.
When one of the neutrons in the projectile collides with a target proton and transforms into a $\Delta$ particle, the remaining nucleons in the projectile recoil.
Following this three-body collision, the generated $\Delta$ particle remains at its original energy level, resulting in a relatively small excitation energy for the projectile-like nucleus.
In the secondary decay process, gamma emission predominates over neutron evaporation. Consequently, the mass number of the final product in the medium recoil mode is larger than that in the cascade scattering mode.

\begin{figure}
\centering
\includegraphics[width=8.6cm,angle=0]{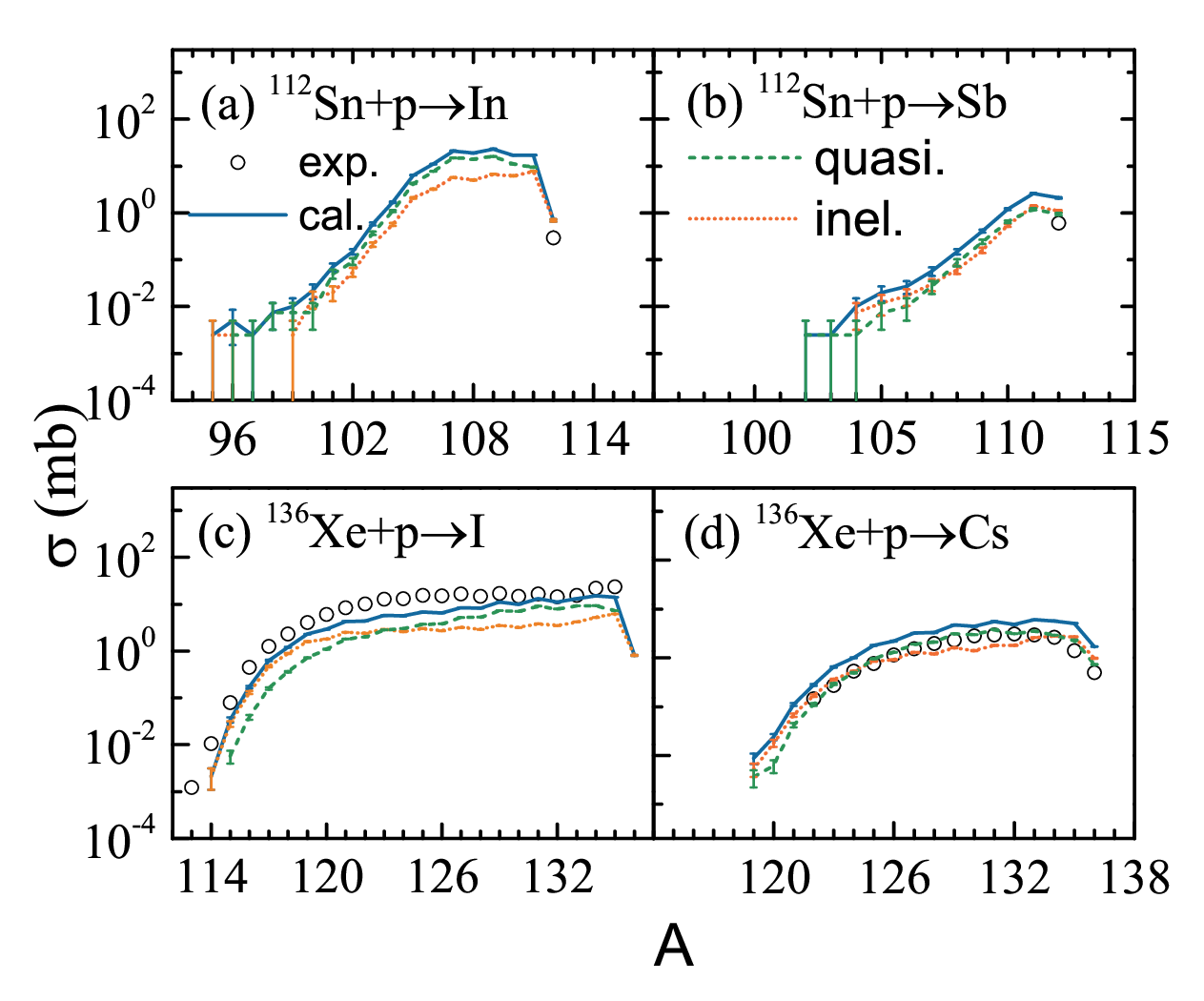}
\caption{Same as Fig. \ref{sigma_cas} but the medium recoil mode is applied in the model.
}
\label{sigma_rec}
\end{figure}

The $^{112}$Sn and $^{136}$Xe + p collisions at 1 GeV/nucleon were simulated using the improved model within the medium recoil mode.
Figure \ref{sigma_rec} displays the production cross sections of In and Sb isotopes from the $^{112}$Sn + p collision, and I and Cs isotopes from the $^{136}$Xe + p collision.
Comparing the calculations with the data reveals a closer agreement, typically within 0.5 orders of magnitude for both collision systems.
For instance, the experimental production cross sections for $^{112}$In and $^{112}$Sb are 0.29$\pm$0.01 and 0.60$\pm$0.02, respectively, whereas the model calculations yield 0.70$\pm$0.04 and 2.10$\pm$0.07.
Although the model still tends to overestimate the cross sections, the discrepancies have notably reduced since adopting the medium recoil mode instead of the cascade scattering mode.

The comparison also highlights that the inelastic contributions, which were significantly enhanced in the medium recoil mode, now exhibit comparable magnitudes to the quasi-elastic contributions.
This improvement underscores the critical role of incorporating the medium recoil mode for accurately modeling single charge-exchange reactions.
Moving forward, the model will continue to utilize the medium recoil mode in subsequent calculations.

\subsection{Single charge-exchange isobaric reaction and pion production}

\begin{table}[htbp]
  \centering
  \caption{Single isobaric charge-exchange reactions and their cross sections in the $^{112}$Sn, $^{136}$Xe + p collision at 1 GeV/nucleon. }
    \begin{tabular}{cllll}
		\hline \hline
		\cmidrule{3-5}    \multicolumn{1}{c}{\multirow{2}[2]{*}{collision}} & \multicolumn{1}{c}{\multirow{2}[2]{*}{production}} & \multirow{2}[2]{*}{channel} & \multicolumn{2}{l}{$\sigma$ ($\mu$b)} \\
          &       &       & cal.  & exp. \\
    \hline
    \multirow{4}[2]{*}{$^{112}$Sn+p} & $^{112}$In+p+$\pi^{+}$ & inelastic &700$\pm$40      & 291$\pm$9 \\
          & $^{112}$Sb+n & quasi-elastic & 940$\pm$50     & 235$\pm$10 \\
          & $^{112}$Sb+p+$\pi^{-}$ & \multirow{2}[1]{*}{inelastic} & \multirow{2}[1]{*}{1100$\pm$50} & \multirow{2}[1]{*}{369$\pm$15} \\
          & $^{112}$Sb+n+$\pi^{0}$ &       &       &  \\
    \hline
    \multirow{4}[2]{*}{$^{136}$Xe+p} & $^{136}$I+p+$\pi^{+}$ & inelastic & 810$\pm$20     &  \\
          & $^{136}$Cs+n & quasi-elastic & 730$\pm$20     &  \\
          & $^{136}$Cs+p+$\pi^{-}$ & \multirow{2}[1]{*}{inelastic} & \multirow{2}[1]{*}{980$\pm$20} & \multirow{2}[1]{*}{} \\
          & $^{136}$Cs+n+$\pi^{0}$ &       &       &  \\
	\hline\hline
    \end{tabular}%
  \label{tab:addlabel}%
\end{table}%

The single isobaric charge-exchange (SICE) reaction is a specialized process within single charge-exchange reactions where the charge number of the projectile-like production changes by one unit compared to the projectile, while the mass number remains unchanged.
In the $^{112}$Sn and $^{136}$Xe + p collisions at 1 GeV/nucleon, the reaction channels of SICE and their corresponding cross sections are summarized in Table \ref{tab:addlabel}.

In the $^{112}$Sn + p collision, the SICE process produces the isobar $^{112}$In or $^{112}$Sb.
For $^{112}$In, only the inelastic channel contributes. The calculated cross section is 700$\pm$40 $\mu$b, approximately 2.5 times larger than the experimental value of 291$\pm$9 $\mu$b.
For $^{112}$Sb, both quasi-elastic and inelastic channels contribute to its production. The calculated cross sections are also higher than the experimental values, although the model reproduces the ratio between the quasi-elastic and inelastic contributions.
In the case of the $^{136}$Xe + p collision, experimental data has not been reported. The calculations indicate that both quasi-elastic and inelastic channels contribute comparably to the production of the isobar $^{136}$Cs, with cross sections of similar magnitudes.

\begin{figure}
\centering
\includegraphics[width=8.6cm,angle=0]{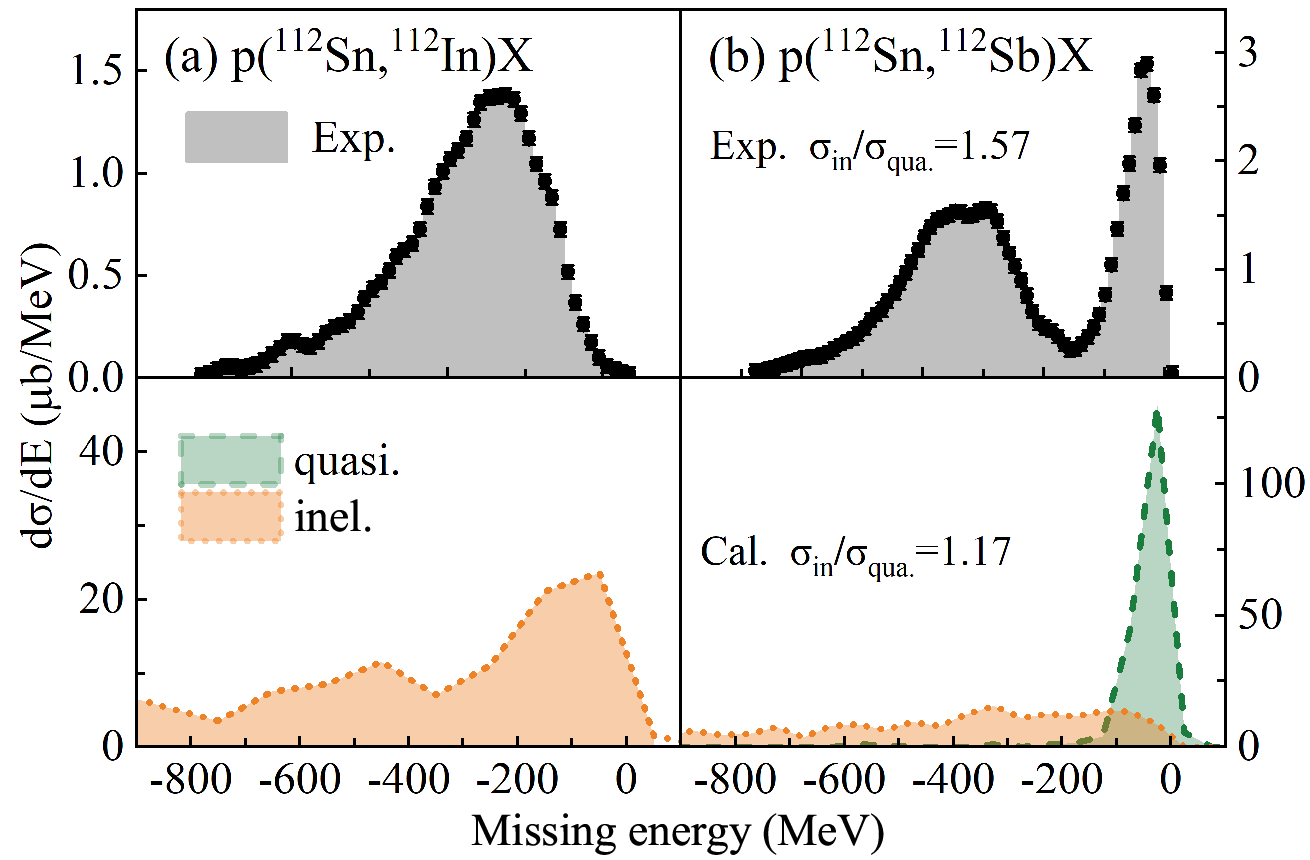}
\caption{\label{miss}
Missing-energy spectra of $^{112}$In and $^{112}$Sb produced in the $^{112}$Sn + p collision at 1 GeV/nucleon.
Grey shaded area are experimental data taken from Ref. \cite{rodriguez2022systematic}.
Orange (Green) shaded area bounded by dotted (dashed) curve represents the missing-energy spectrar of inelastic (quasi-elastic) channel.
Left (Right) panel is the missing-energy spectra of $^{112}$In ($^{112}$Sb).
Ratios of inelastic to quasi-elastic contributions for experimental data and calculations are presented.}
\end{figure}

A technique utilizing missing-energy spectra has been applied to distinguish the quasi-elastic and inelastic components in the single isobaric charge-exchange (SICE) reaction, as reported in Ref. \cite{rodriguez2022systematic}.
Missing energy refers to the kinetic energy difference between the projectile-like production and the initial projectile, serving as a measure of energy dissipation during the SICE process into the projectile.

As depicted in Figs. \ref{miss}(a) and (b), the experimental missing-energy spectrum shows a broad peak for $^{112}$In production and features two distinct components for $^{112}$Sb production.
In the inelastic channel, lower missing energies correspond to energy dissipation where a nucleon is excited into a $\Delta$ particle within the projectile nuclear system. Conversely, quasi-free charge-exchange processes in the quasi-elastic channel result in a near-zero missing energy.
The two components observed for $^{112}$Sb production effectively distinguish between quasi-elastic and inelastic events.
The experimental inelastic cross section is 369$\pm$15 $\mu$b, while the quasi-elastic cross section is 235$\pm$10 $\mu$b, resulting in an inelastic-to-quasi-elastic ratio of approximately 1.57.
The calculations performed by the IQMD+GEMINI++ model, illustrated in Figs. \ref{miss}(c) and (d), successfully reproduce the overall characteristics of the experimental spectra.
For $^{112}$Sb production, the ratio between the inelastic and quasi-elastic components is 1.17, which closely matches the experimental value of 1.57.
This underscores the crucial role of incorporating the medium recoil mode to accurately capture the dynamics of single charge-exchange reactions.
However, the model fails to accurately reproduce the peak position and width of the inelastic component.
Improvements in these aspects are necessary for future advancements of the model.

\begin{figure}
\centering
\includegraphics[width=8.6cm,angle=0]{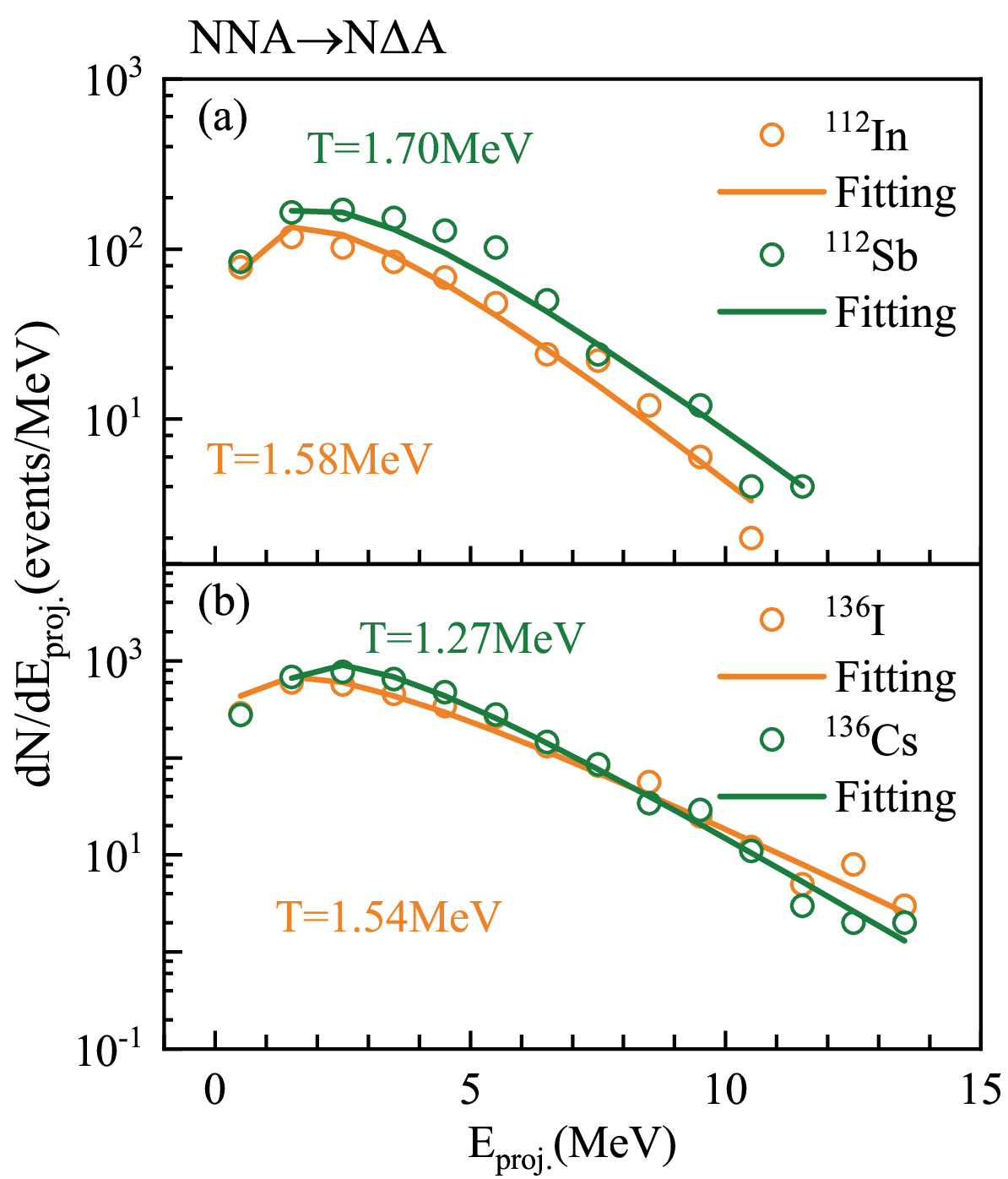}
\caption{\label{Ek_pion}
Kinetic energy spectra of pions associated with the production of (a) $^{112}$In and $^{112}$Sb, and (b) $^{136}$I and $^{136}$Cs produced in the SICE reaction.
}
\end{figure}

The kinetic energy spectra of pions associated with the production of $^{112}$In and $^{112}$Sb produced in $^{112}$Sn + p collision and $^{136}$I and $^{136}$Cs produced in the $^{136}$Xe + p collision are predicted and shown in Fig. \ref{Ek_pion}.
The spectra have been transformed to the center-of-mass frame of the projectile, which can measure the thermalization of the nuclear system.
It is shown that the kinetic energy distributions in both collisions are similar.
The most probable kinetic energy is 2.5 MeV, and the energies for most events are less than 15 MeV.
The slope temperature are extracted by the Maxwellian fitting of the spectra,
\begin{equation}
Y(E)\propto\frac{E-E_0}{T_{\mathrm{slope}}^2}\exp\left(-\frac{E-E_0}{T_{\mathrm{slope}}}\right),
\end{equation}
where the E${_0}$ is stand for the effect of Coulomb repulsion, and the T$_{\mathrm{slope}}$ is the slope temperature.
In the $^{112}$Sn + p collision, the slope temperature of $^{112}$In and $^{112}$Sb is 1.58 MeV and 1.70 MeV, respectively.
The value is 1.54 MeV for $^{136}$I and 1.27 MeV for $^{136}$Cs in the $^{136}$Xe + p collision.

\begin{figure}
\centering
\includegraphics[width=8.6cm,angle=0]{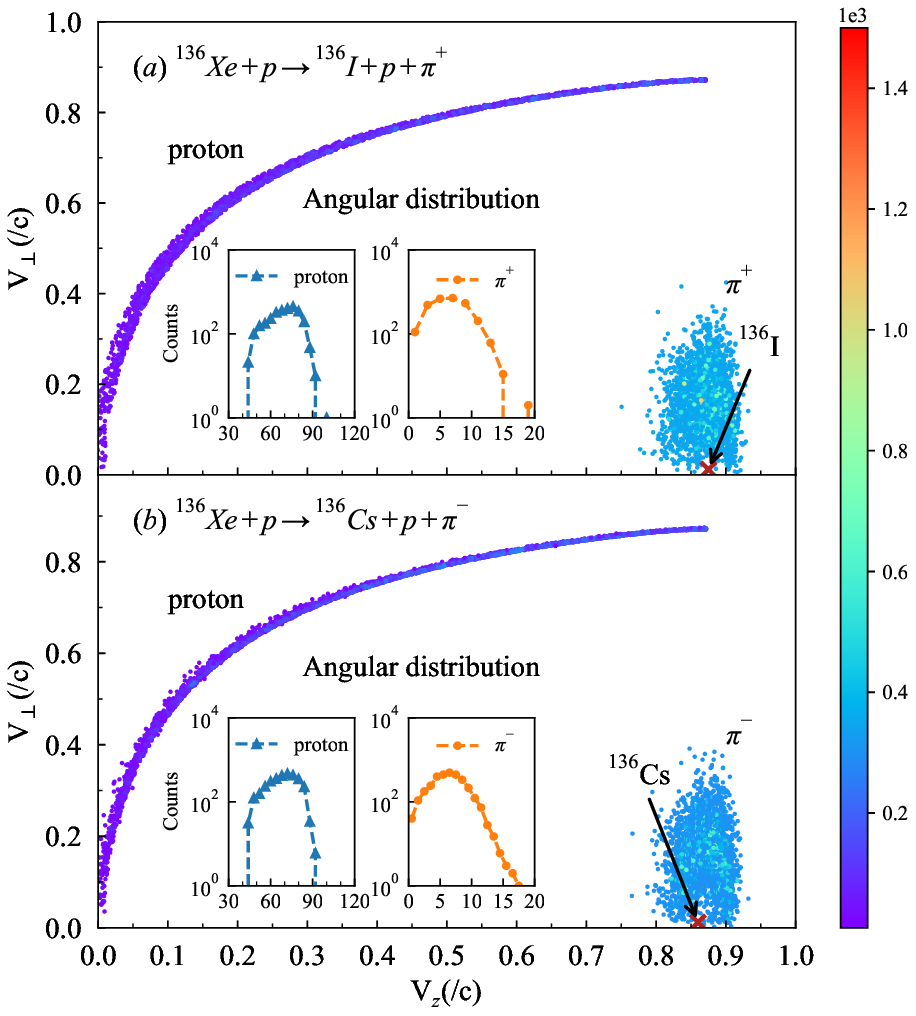}
\caption{\label{Angle_pion}Correlation of longitudinal and transverse velocity of protons and pions produced in (a) $^{136}$Xe + p $\to$ $^{136}$I + p + $\pi^{+}$ at 1 GeV/nucleon, and (b) $^{136}$Xe + p $\to$ $^{136}$Cs + p + $\pi^{-}$. The angular distributions are also provided as the insert figure.}
\end{figure}

To further investigate the $\Delta$ dynamics in upcoming experiments, Fig. \ref{Angle_pion} illustrates the plane of transverse velocity versus longitudinal velocity of protons and charged pions produced in single isobaric charge-exchange (SICE) reactions during the $^{136}$Xe + p collision at 1 GeV/nucleon.
The angular distributions are detailed in the embedded figures.

Panels (a) and (b) depict the reaction channels associated with the production of $^{136}$I and $^{136}$Cs, respectively.
Both SICE productions are observed to be closely aligned with the incident direction of the projectile $^{136}$Xe, within an angular range of approximately 1$^{\circ}$.
Given their identical mass number and a small difference of only two units in atomic number, the velocity and emission angular distributions of protons and charged pions produced with them exhibit remarkable similarities.

In the figure, the longitudinal and transverse velocity distributions of protons exhibit a wide dispersion, ranging from 0 to 0.9 $c$.
The velocity distribution shows slight fluctuations around a curve.
Proton emission angles are concentrated prominently within the range of 45$^{\circ}$ to 90$^{\circ}$, peaking around 75$^{\circ}$, indicating a higher probability of protons being emitted in the longitudinal direction. Detectors strategically placed within the range of 45$^{\circ}$ to 90$^{\circ}$ relative to the incident direction can maximize the experimental detection efficiency of protons associated with $^{136}$I.

In comparison to protons, the distribution of pions is more dispersed.
In panel (a), the majority exhibit transverse velocities below 0.4 $c$, while their longitudinal velocities are predominantly observed within the range of 0.8 $c$ to 0.95 $c$.
The angular distribution of pions shows a notable concentration of emission angles within the range of 0$^{\circ}$ to 20$^{\circ}$, with the peak occurring at approximately 7.5$^{\circ}$.
Experimental detectors positioned within this angular range can effectively resolve events, resulting in a relatively high detection efficiency of pions.

\section{\label{summary}Summary}
In summary, the dynamic mechanisms underlying single charge-exchange reactions have been investigated using a theoretical framework that combines the Isospin-dependent Quantum Molecular Dynamics model with the statistical decay model GEMINI++.
Two distinct channels contribute to the single isobaric charge-exchange (SICE) reaction: quasi-elastic channels, where neutron-proton scattering drives the charge-exchange, and inelastic channels, where the $\Delta$ resonance is produced during the collision.
In Ref. \cite{rodriguez2022systematic}, a technique based on the missing-energy spectra is employed to distinguish the quasi-elastic and inelastic contributions of SIEC cross sections.
It is reported that the inelastic cross section is 369 $\pm$ 15 $\mu$b, and the quasi-elastic cross section is 235 $\pm$ 10 $\mu$b, yielding a ratio of inelastic cross section to quasi-elastic cross section of approximately 1.57.
Considering the novel medium recoil mode associated with $\Delta$ production considered in the model, the calculations of the missing-energy spectra can reproduce the main characteristics of the experimental data successfully, and the ratio between the inelastic cross section to quasi-elastic cross section is 1.17 for $^{112}$Sb in the $^{112}$Sn + p collision.

The mainstream microscopic transport models have yet to incorporate the in-medium effect stemming from inelastic nucleon-nucleon collisions.
The impact of the medium recoil mode on the characteristic features of pions is under discussion.
Velocity spectra of protons and pions in the context of SICE reactions are also predicted, with the aim of investigating $\Delta$ dynamics in upcoming experiments. The emission angle of pions is concentrated within approximately 20 degrees of the incident direction.
Measurement of the pions produced in SICE reactions will provide insights into the in-medium effects of $\Delta$ dynamics, adding a valuable dimension to our understanding of the complex dynamics involved in these reactions.

\section*{ACKNOWLEDGMENTS}

This work was supported by the National Natural Science Foundation of China under Grants Nos. 12475136
and 12075327, the Central Government Guidance Funds for Local Scientific and Technological Development, China (No. Guike ZY22096024), the Key Laboratory of Nuclear Data foundation (No. JCKY2022201C157).

\bibliography{apssamp_prc}

\end{document}